# Modelling the enhancement of spectrally broadband extraction efficiency of emission from single InAs/InP quantum dots at telecommunication wavelengths

Pawel Mrowiński,[*] Grzegorz Sęk

*Laboratory for Optical Spectroscopy of Nanostructures, Department of Experimental Physics, Faculty of Fundamental Problems of Technology, Wrocław University of Science and Technology, Wybrzeże Wyspiańskiego 27, 50-370 Wrocław, Poland*

\* corresponding author: pawel.mrowinski@pwr.edu.pl

**Abstract**

Hereby, we present results of finite-difference time-domain simulations of technologically undemanding photonic structures offering a significant improvement in the spectrally broad extraction efficiency of emission from a quantum emitter at the third telecommunication window. The modelling is performed for cylindrical and cuboidal mesa structures containing an exciton confined in a single InAs/InP quantum dot. We investigate the emission pattern from such mesas made of InP (which can be exchanged with AlGaInAs lattice-matched to InP) for two cases: monolithic InP material or InP/AlGaInAs distributed Bragg reflector (DBR) underneath the mesa. The calculations are performed as a function of geometrical parameters, i.e. mesa shape and dimensions, and the emitter position. For the case with the DBR, we obtain extraction efficiencies above 25% (for a numerical aperture of 0.4 of a common optical detection system), which also exhibits weak spectral dependence – values above 20% can be obtained in the range of 60 nm around 1550 nm. When the numerical aperture is increased to 0.9, extraction efficiencies about

45% are reachable within the proposed solution. The extraction efficiency is rather weakly sensitive to the in-plane dipole alignment (in the range of 100-200 nm), however, it is more susceptible to even slight changes in its vertical position or to mesa sizes. Eventually, we also comment on the Purcell effect in such weak photonic confinement structures, influence of which on the source brightness is not negligible in some cases.

## 1. Introduction

A single quantum dot (QD) is an example of a quantum emitter, which can be created in a semiconductor matrix and used for generation of single photons or entangled photon pairs on demand [1–3], or even more complex multiphoton states [4,5]. The maximum single photon count rates detected externally for such emitters are mainly limited by the light collection efficiency. In the case of standard homogeneous medium, when QD emission is isotropic, the fraction of collected signal is at best on the level of a few percent, which is related to the limited numerical aperture (NA) of detection systems or transmission channels and to the intrinsic effect of total internal reflection at the semiconductor-air interface.

The issue of small efficiency of photon extraction can be overcome by engineering the photonic confinement in nano- or microstructures, which can induce directionality of emission due to cavity effects, i.e. by coupling with cavity modes that leak out in a specific direction, or due to engineering of refraction by shaping the semiconductor-air interface. There exist at least several approaches, including 1D microcavities with distributed Bragg reflectors (DBRs) [6,7], 2D photonic crystal nanocavities [8], or circular dielectric gratings [9,10]. However, in spite of high performance in the listed demonstrations, they usually suffer from a narrow spectral band of operation related to cavity mode linewidths, typically on the order of sub-nanometer to a few nanometers, and therefore it is often necessary to tune emitters spectrally with high precision by using external triggers. This makes such approaches rather impractical in the sense of large volume fabrication of devices with repeatable characteristics. Spectrally broad extraction efficiency functions can be obtained using photonic nanowires [11], microlens structures [12], solid immersion lenses [13] or the easiest to fabricate mesa structures [14,15] . However, these have mostly been reported for the structures made in GaAs-based material system and emitting below 1 µm. When concerning the telecommunication range of 1.3 or 1.55 µm suitable for quantum communication protocols in short or long haul fiber networks, the InP-based material system is mainly considered [16–21]. In that case, fabrication techniques are less mature and optimization in terms of high extraction efficiency for subsequent effective outcoupling is still an ongoing research. There only exist a very few reports on spectrally narrow extraction efficiency in InP-based microcavity systems at 1.3 µm [22], whereas the highest obtained extraction efficiencies for spectrally broad approaches are just exceeding 10% at both 1.3 µm (on GaAs) [23,24] as well as 1.55 µm (on InP) [25].

In this work we show, by numerical simulations, that a photonic structure exhibiting high and spectrally broad extraction efficiency at 1.55 µm can be obtained in an InP-based material family by using a technologically simple mesa structure of cylindrical or cuboidal geometry with a DBR beneath the mesa/emitter, in analogy to GaAs-based structures operating at shorter wavelengths [12,14], [23]. We model the InP-based system, which provides the flexibility of operation tunability through the 2nd and 3rd telecom windows.

We focus on optimization leading to a collection of a significant fraction of a dipole source emission along the growth direction (z-axis). In order to optimize the extraction efficiency, we scan over parameter space related to the mesa geometry and dipole position. We evaluate the extraction efficiency ($\eta_{eff}$) for normalized dipole power, concentrating on directionality due to the photonic structure and we analyze NA dependence to find the $\eta_{eff}$ limits related to a hypothetical "first lens" in an optical detection system (i.e. a closely spaced fiber or a microscope objective). Finally, the influence of dielectric environment and the related optical confinement on the local density of states (LDOS) is analyzed by means of the Purcell factor ($F_p$) in order to deduce on the highest possible brightness of the photonic mesa structure in terms of single photons or entangled photon pairs emission utilizing embedded QD emitter. The Purcell effect included in our approach, which is usually ignored in other works on similar photonic engineering, can offer additional benefits for optimizing the source brightness.

## 2. Investigated structures

Recently, we have demonstrated optimization of the photonic environment to tailor the polarization anisotropy of emission from a single quantum emitter of an InAs/AlGaInAs/InP material system [18]. Hereby, we focus on the improvement of extraction efficiency of the photonic structures in a form of cylindrical (sec. 4.2) or cuboidal (sec. 4.3) mesas with an embedded dipole emitter, which imitates a quantum dot bright exciton state. The mesa is made of InP or compatible $Al_{0.24}Ga_{0.23}In_{0.53}As$ material, lattice matched to InP [16,21,26–29], which can be interchanged in the fabrication of nanophotonic devices aimed at the telecom range due to similar refractive index, therefore our approach is valid for both of them. The mesa is placed on monolithic

InP - sample "A" - or a DBR on InP - sample "B" - as shown in Fig. 1. The DBR structure optimized for 1.55 μm wavelength is made of InP (122.4 nm) and $Al_{0.05}Ga_{0.42}In_{0.53}As$ (109.5 nm), with the refractive indices of 3.17 and 3.54, respectively [30,31].

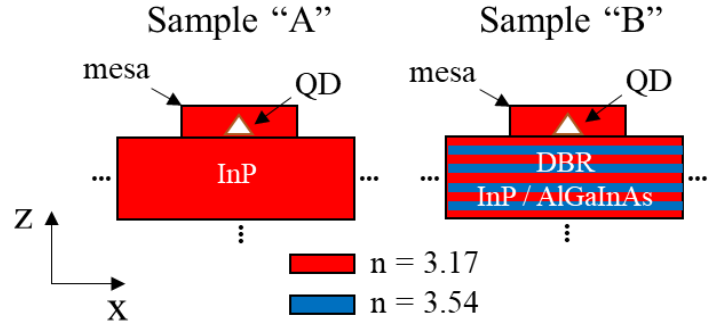

Figure 1: Schematic view of photonic mesa structures with embedded QD simulated by a linear dipole (white triangle) on InP substrate (sample "A") and on 16 pairs of InP/AlGaInAs DBR mirror on InP (sample "B").

### 3. Finite-difference time-domain model

We use the finite-difference time-domain (FDTD) commercial-grade solver of Maxwell's equations (Lumerical software [32]) in order to perform calculations of electromagnetic field generated by a point dipole emitter embedded in a photonic mesa structure. A schematic view of the 3D FDTD model is presented in Fig. 2(a), where within the computational domain we have the photonic structure, the single optical dipole and two near-field monitors, first 1.65 μm above the substrate (size of 3 x 3 μm²) and second in the vertical cross-section (size of 6 x 3 μm²). The computational domain is defined as a 4 x 4 x 8 μm³ cuboid and it is limited by the perfectly matched layer boundary conditions. Within this model, we calculate the far-field projection for the hemisphere above the mesa structure, using the upper near-field monitor. The far-field projection of the generated electric field is then used for analysis of the emission cone, which allows characterizing the fraction of a dipole emission from the mesa structure along the z-direction that corresponds to a typical experimental configuration used for collection of the quantum dot emission in spectroscopic experiments and for single-photon emission characterization. The 2D cross-sectional near-field is used for visualizing the electric field intensity distribution inside the mesa and in the mesa proximity. Exemplary results for the calculated far-field presented on the upper hemisphere and the near-field in the cross-section using the presented model can be seen in Fig. 2(b), for both "A" and "B" samples.

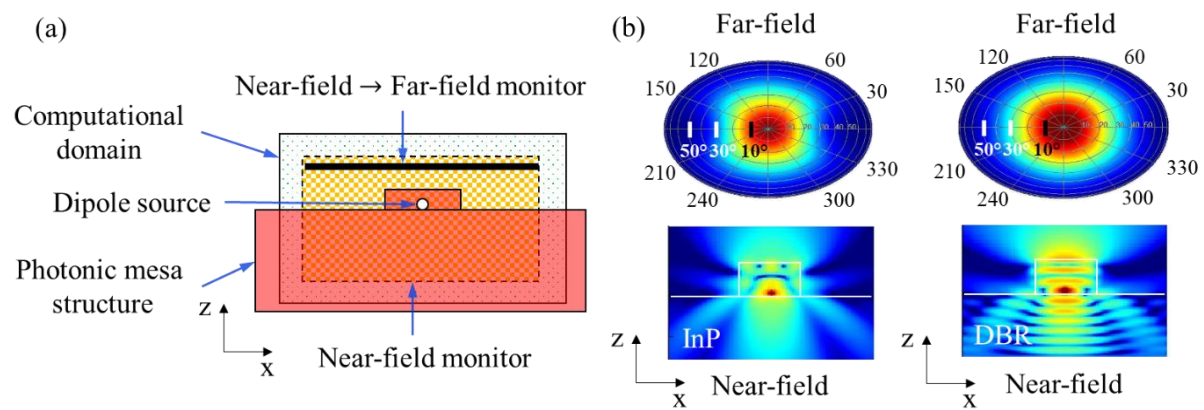

Figure 2: (a) Schematic view of the 3D FDTD model structure used in the calculations of extraction efficiency. The computational domain consists of the photonic mesa structure with embedded linear dipole, cross-sectional near field monitor plane and upper near-field monitor serving for the far-field analysis. (b) Exemplary results for two samples "A" (left) and "B" (right) from Fig. 1 showing both the far-field $|E|^2$ distribution on the upper hemisphere and the near-field $|E|^2$ distribution in the cross-section generated by the dipole embedded in the photonic mesa structures. The dipole position was set to 50 nm above the substrate, the mesa height is 800 nm and diameter is 900 nm. Both images are presented in the same logarithmic color scale.

## 3.1. Extraction efficiency

A standard definition of the extraction efficiency is given by the ratio of the collected fraction of the dipole emission ($P_{NA}$) to the total emitted power ($P_d$):

$$\eta_{eff} = \frac{P_{NA}}{P_d}, \qquad (1)$$

The collected fraction of emission, which is related to the light cone propagating along the z-direction, is then defined by:

$$P_{NA} = \frac{1}{2} \int Re(\vec{P}) d\vec{S} \times I_{ff,NA}, \qquad (2)$$

where $\vec{P}$ is the Poynting vector normal to the surface above the mesa structure [region of the near-field → far field monitor in Fig. 2(a)], $d\vec{S}$ is the surface normal vector and $I_{ff,NA}$ is the fraction of the integrated far-field intensity within the analyzed numerical aperture. The power calculated in eq. (2) is related to the dipole emission influenced by the LDOS of the photonic structure (due to Purcell effect), however, this influence is in fact eliminated from $\eta_{eff}$ as defined in (1), by taking into account that the dipole power $P_d = F_p \cdot P_S$, where $F_p$ is the Purcell factor ($F_p > 1$ – enhancement, $0 < F_p < 1$ – inhibition), and $P_S$ is the intrinsic source power, i.e. when the dipole is placed in the homogeneous medium or vacuum. Therefore, the $\eta_{eff}$ in (1) quantifies only the directionality of emission due to refraction or reflection of light.

All the necessary components for the evaluation of $\eta_{eff}$ are obtained from the FDTD simulation results according to the description in ref. [33]. The integrated far field emission intensity within the selected NA ($I_{ff,NA}$) over the upper detection plane above the mesa structure is given by:

$$I_{ff,NA} = \frac{\int_0^{\varphi=2\pi} \int_0^{\theta_{HA}} E^2(\theta,\varphi) \sin(\theta) d\theta d\varphi}{I_{ff}}, \quad (3)$$

where, $E^2(\theta,\varphi)$ is the electric field intensity, $\theta_{HA}$ and $\varphi$ are the polar half-angle and the azimuth angle in spherical coordinates system, respectively, and $I_{ff}$ is the total integrated far field intensity. For instance, the NA of 0.4 corresponds to the light cone limited by the half-angle of $\theta_{HA} \cong 23.58°$.

## 4. Results

### 4.1. Influence of DBR on extraction efficiency

In order to optimize the extraction efficiency we calculate first the reflectivity of the DBR structure. The spectral dependence could be obtained by using the 3D FDTD simulation of a reflected plane wave for various wavelengths. In Fig. 3, we show the reflectivity and the transmission spectra in cases of 16 and 32 mirror pairs, resulting in around 83% and 95% of reflection at 1.55 μm, respectively, where the stopband is spectrally expanded from about 1500 to 1600 nm, which is favourable for a broadband operation. Such FDTD evaluation of the DBR reflection gives more realistic results than any simplified analytical reflectivity spectrum calculations [34], as in the used approach it includes non-trivial scattering or absorption and the related losses $(R + T < 1)$, which might also be relevant in the context of extraction efficiency.

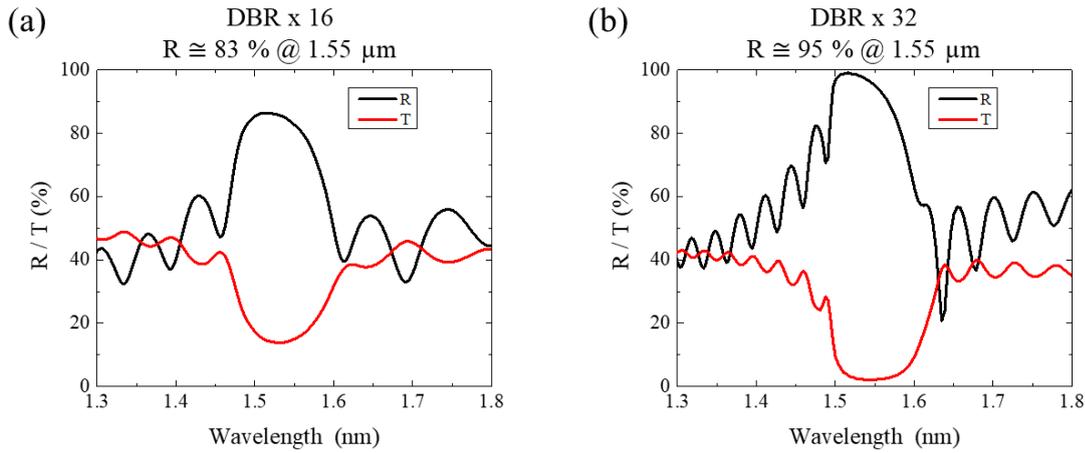

Figure 3. Reflection and transmission of DBR planar mirror composed of (a) 16 and (b) 32 InP/AlGaInAs pairs calculated in 3D FDTD.

We derive the effect of the DBR on the extraction efficiency for an emitting dipole embedded in an exemplary planar InP structure of 800 nm height, and for the dipole position 250 nm above the DBR (these are just exemplary system parameters). In Fig. 4, we present the calculated extraction efficiency as a function of the number of InP/AlGaInAs bilayers, from 0 to 32 pairs. We used several numerical apertures of 0.2, 0.4, 0.6 and 0.8 to simulate various optical detection systems. For the case without the DBR, we deal with a monolithic InP substrate and the values of extraction efficiency are 0.2%, 0.6%, 1.2%, 1.7%, respectively. Introducing DBR structure of only several mirror pairs results already in a significant increase in $\eta_{eff}$ due to enhanced back reflection of the dipole emitted light. One can observe a saturation of $\eta_{eff}$ for about 12-20 DBR layers' pairs, depending on the NA. It shows a good trade-off between technological efforts of the DBR fabrication and the resulting extraction efficiency, maximum of which can be reached for a reasonable number of mirror pairs.

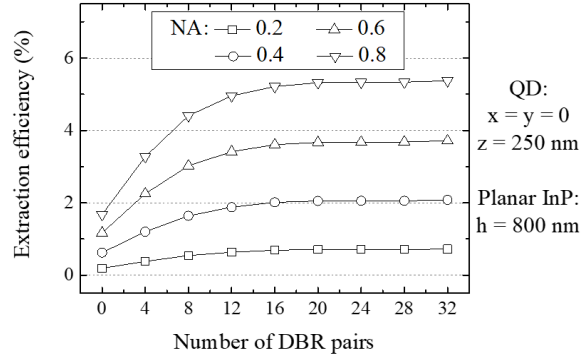

Figure 4. Extraction efficiency for planar InP structure for several NA from 0.2 to 0.8 calculated as a function of the number of DBR layers' pairs.

*4.2. Circular mesa structure*

Next, we study the extraction efficiency for the numerical aperture of 0.4 for the following sequence of parameters: dipole vertical z-position, dipole in-plane x-position, mesa height and mesa diameter. While it is a 4-dimensional space of parameters and the computational effort required to study the high-resolution grid is high, we perform first a set of preliminary calculations for a small-resolution grid to establish fixed remaining parameters for a single series. Preliminary results are collected by the calculations within the following parameter space: I. mesa width (or diameter) of 400:100:1000 nm (start:step:stop), II. mesa height 400:100:900 nm, III. QD vertical position 0:50:300 nm, IV. QD in-plane position in the center. The strategy of parameters' selection in the following series of calculations is associated with the best outcome in terms of the extraction efficiency and Purcell factor. Then, we calculate single scans for every parameter, with the other parameters being fixed. As from preliminary data set, we find extreme outcomes (when looking at the extraction efficiency mainly, but also at the Purcell factor) for mesa geometry of about 900 nm width (or diameter) and about 800 nm height, for several QD vertical positions, so we fix them for all the series presented in Figs 5 and 6. In the case of Fig. 5(b) and Fig. 6(b), we choose $z$ = 50 nm, as it can be seen in Fig. 5 (a) that it provides $\eta_{eff}$ >10% and $F_p$ > 1.5 for the sample "B" or based on Fig. 6(a) it gives $\eta_{eff}$ >10% and $F_p$ > 1.0, which we find useful to further examine the detuning of the in-plane position from the mesa centre to demonstrate the relative decrease of the extraction efficiency. Therefore, we also keep $z$ = 50 nm in the case of tuning the mesa height in Figs. 5(c) and 6(c). However, for Figs. 5(d) and 6(d) we use another "good" vertical QD position (for $z$ = 250 nm), which is close to the best value in terms of extraction efficiency (and still $F_p \geq 1$), as seen in Figs. 5(a) and 6(a). We are aware that such a methodology cannot assure pointing out the exactly maximal absolute values, but within a reasonable computational effort it was a kind of compromise to show selected examples of the high efficiency in the case of such mesa structures. We believe that it allowed getting close to the optimal values, although they could differ slightly when, for instance, a higher resolution grid is used. Moreover, our additional goal regarding Fig. 5 and Fig. 6 is to present clearly the trends related to the tuning of a specific parameter and the interrelation between the extraction efficiency and the Purcell factor, which are often omitted – see the discussion further below.

First, we discuss the dependence on the dipole position along the *z*-direction, for the in-plane position fixed in the centre ($x = y$ = 0 nm). The mesa geometry is chosen to be 900 nm in diameter and 800 nm in height, following the preliminary results on the low resolution grid. The results for both InP substrate – sample "A" – and DBR mirror (16 pairs) – sample "B" – are presented in Fig. 5(a) - solid and open points, respectively. We observe a rather flat dependence for the sample "A" in the range of $z$ = 0÷450 nm of $\eta_{eff}$

~ 10%, and above $z$ = 450 nm the $\eta_{eff}$ drops down considerably, due to proximity of the upper interface, causing scattering in the horizontal direction. In the case of the sample "B", we find an oscillating behaviour with the extreme values of $\eta_{eff}$ = 19.1% and $\eta_{eff}$ = 1.6% at $z$ = 225 nm and $z$ = 400 nm, respectively. The period of oscillations indicates the optical cavity effect along the $z$-direction introduced by the DBR and the InP/air interface, where ~3(4) nodes (antinodes) of the electric field intensity can be found for 1.55 µm wavelength [cf. Fig. 2(b) – sample "B"].

In Fig. 5(b) we examine the influence of the in-plane dipole position change up to 300 nm from the centre of the mesa. We use the same mesa geometry and the dipole $z$-position of 50 nm to investigate the effect of the in-plane dipole displacement. As could be expected, we observe here a smooth decrease of the $\eta_{eff}$ in both samples reaching 0.7% and 2.3% at $x$ = 300 nm. We notice also that the gradient is rather small up to about $x$ = 125 nm and $x$ = 150 nm for the samples "A" and "B", respectively, which gives only a relative decrease of about 25% (2.0% and 3.9% of absolute decrease), and so, the in-plane positioning accuracy of the QD within ±50 nm range would be satisfactory for realization of such an optimized device [23].

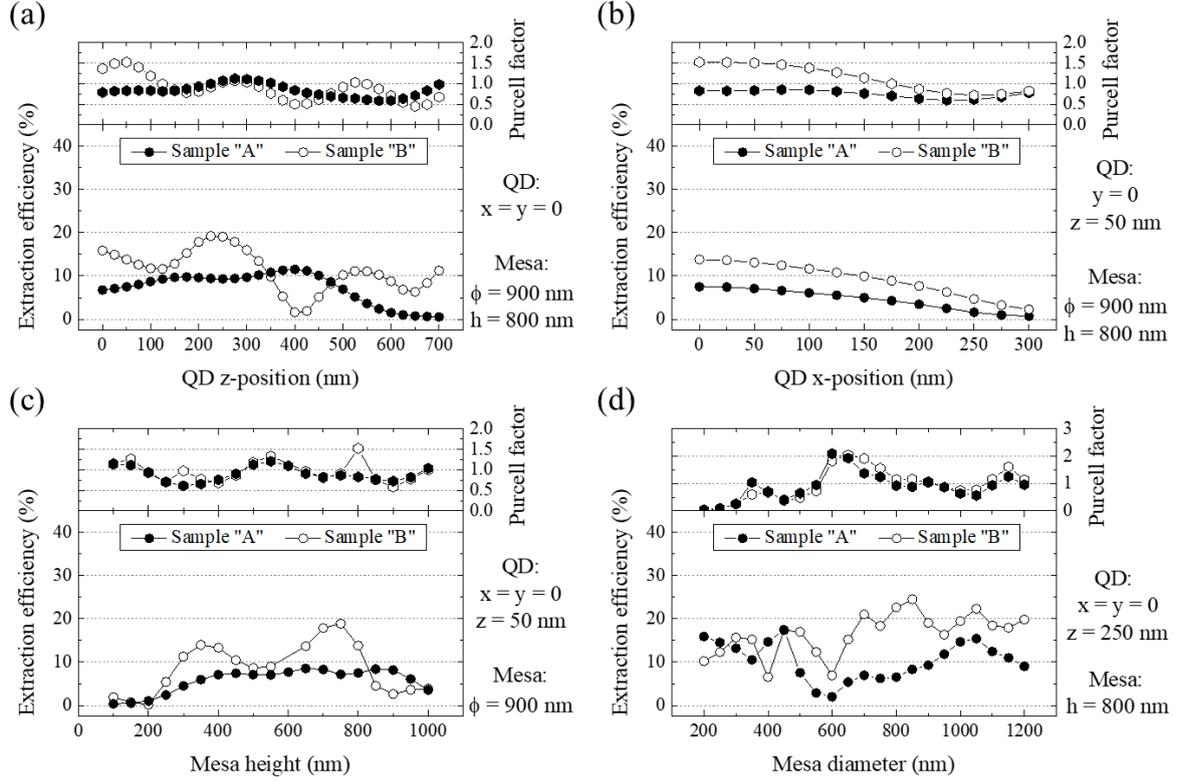

Figure 5. Calculated extraction efficiency and Purcell factor in dependence of the QD position in the photonic mesa structure along vertical z-direction (a) and in-plane x-direction (b), as well as their dependence on the mesa height (c) and diameter (d). The fixed parameters (see text) are shown on the right side of the graphs.

Furthermore, we investigate the extraction efficiency dependence versus parameters related to the mesa geometry, i.e. the mesa height and diameter. In Fig. 5(c) we use the dipole positioned at the mesa centre and $z$ = 50 nm above the InP/DBR section, and we tune the mesa height from 100 nm to 1000 nm, while having the diameter fixed at 900 nm. In the case of the sample "A", we find a smooth increase of $\eta_{eff}$ up to 8.5% at 650 nm height and its decrease above 900 nm height down to $\eta_{eff}$ = 3.6%. The characteristics of the sample "B" exhibit oscillations with the maxima at 350 nm and 750 nm of the mesa height ($\eta_{eff}$ = 14% and $\eta_{eff}$ = 18.9%), respectively. Such fluctuations are also related to the overlap between the antinodes of the electric field intensity with the dipole *z*-position due to the cavity effect along the *z*-direction.

In Fig. 5(d), we use the mesa height of 800 nm, dipole *z*-position of 250 nm and we change the mesa diameter from 200 nm to 1200 nm. In both the cases of "A" and "B" samples, we see rather irregular dependence and abrupt changes in $\eta_{eff}$, being related to the in-plane optical confinement of the mesa structure. The "A" sample shows the maximum of $\eta_{eff}$ = 17.5% at $\phi$ = 450 nm, while the "B" sample has the maximum of $\eta_{eff}$ = 24.6% at $\phi$ = 850 nm.

As mentioned above, the extraction efficiency defined in (1) expresses only the directionality effect caused by the photonic structure. Such method is reasonable when the quality factor of our photonic structures is very small and thus one could expect $F_P$ ~ 1 in the very first approximation, which on the other hand would be beneficial for achieving a broadband device and high extraction along the *z*-direction. However, we find the Purcell factor noticeably changing in the range from 0.5 to 2.1, which should not be fully neglected (top semiconductor/air interface can provide reflection coefficient of 40% and the bottom DBR x16 mirror approx. 83%), and therefore, we include in Fig. 5 the Purcell factor dependence in the upper panels.

In order to combine the effects of both, the extraction efficiency and the Purcell factor, into one figure of merit, one can employ the so called brightness (see e.g. N. Somaschi et al. [35]), in analogy to single photon sources based on microcavity structures. In the simplest approximation it can be defined by β·$\eta_{eff}$, assuming that the probability of the QD to be excited into the desired (emitting) state upon an excitation pulse equals 1. The parameter *β* = $F_p/(F_p+1)$ expresses the fraction of the emission coupled into the detected mode. Although the absolute values of the Purcell factor are not very significant here (see Fig. 5), they still correspond to changes in $\beta$ from very small values to above 0.6. Moreover, the $F_p$ and $\eta_{eff}$ reach their maximal values for slightly different system parameters, therefore, optimization of the source brightness to that extent requires considering both, the extraction efficiency and the Purcell effect. For example, when looking at Fig. 5(c), we see that for the sample "B", by taking into account the case of $F_P$ = 1.52 and $\eta_{eff}$ = 13.8% for the mesa height of 800 nm, one can obtain the brightness comparable to the mesa of 750 nm height with greater $\eta_{eff}$ = 18.9% and lower $F_P$ = 0.91. In Fig. 5(d), we find the maximum brightness for the mesa diameter of 700 nm, for which the $\eta_{eff}$ = 21.1% and $F_P$ = 1.91 instead of 850 nm mesa with $\eta_{eff}$ = 24.6% and $F_P$ = 1.16. These individual examples confirm that focusing on the directionality or Purcell effect exclusively may be misleading and indicate lower performance of the photonic device than when taking into account their combined effect.

### 4.3. Cuboidal mesa structure

Analogical analysis is performed for a square base cuboidal mesa structures, which are often processed for single quantum dot spectroscopy [18]. In Fig. 6(a,b), there are shown the results of the calculated extraction efficiency for NA = 0.4 and the Purcell factor for both the "A" and "B" samples as a function of the dipole position along the $z$-direction and along the in-plane $x$-direction, respectively. The position change along the z-direction allows obtaining the extraction efficiency on the level of 15.2% (15.7%) for the "A" ("B") sample with $z$ = 200 nm ($z$ = 250 nm) with the $F_p$ = 0.6 ($F_p$ = 0.94). In the case of the sample "A", further increase in the QD position in the $z$-direction results in a decrease of the $\eta_{eff}$, however, for the sample "B", the cavity effect is seen as an oscillating behaviour, both in terms of the $\eta_{eff}$ and the Purcell factor, where the latter is the largest at $z$ = 500 nm ($F_p$ = 2.40). In this case we should also note that due to changes in $F_p$ the device brightness at $z$ = 500 nm would be comparable to $z$ = 250 nm where the highest $\eta_{eff}$ is observed. As expected, changes in the in-plane position of the emitter result in rather smooth decrease of the $\eta_{eff}$ and the relative decrease of more than 25% is seen at $x$ = 125 nm ($x$ = 225 nm) for the "A" ("B") sample, which is again satisfactory from the point of view of technologically reachable positioning accuracy of around ± 50 nm.

In the next step, we examine similar dependencies on the sizes of the mesa structures, i.e. their height and width, as shown in Fig. 6(c,d). In the case of tuning the mesa height, we observe the largest extraction efficiency at 700 nm for both the "A" and "B" samples, with values of 13.4% and 17.6%, respectively. However, the Purcell factor is less than 0.7 in both the cases, which reduces the brightness for such mesas. We should note that here the $\eta_{eff}$ dependence is qualitatively similar, but reversed to the one observed when tuning the dipole z-position, as shown in Fig. 6(a). And as it was previously, the oscillatory kind of dependence is also seen due to the cavity effect along the z-direction. Finally, in Fig. 6(d) we observe a more irregular behaviour versus the mesa width due to complex in-plane optical confinement changes, influencing both the directionality and LDOS. We can find maximal $\eta_{eff}$ = 24.4% and $F_p$ = 1.77 at 650 nm width of the sample "B", which in total results in the best brightness of the considered photonic structures for NA = 0.4. In the case of the sample "A" the maximum $\eta_{eff}$ = 18.2%, however, the Purcell factor is rather low and equal to $F_p$ = 0.39, making it much less beneficial than for the sample with a DBR.

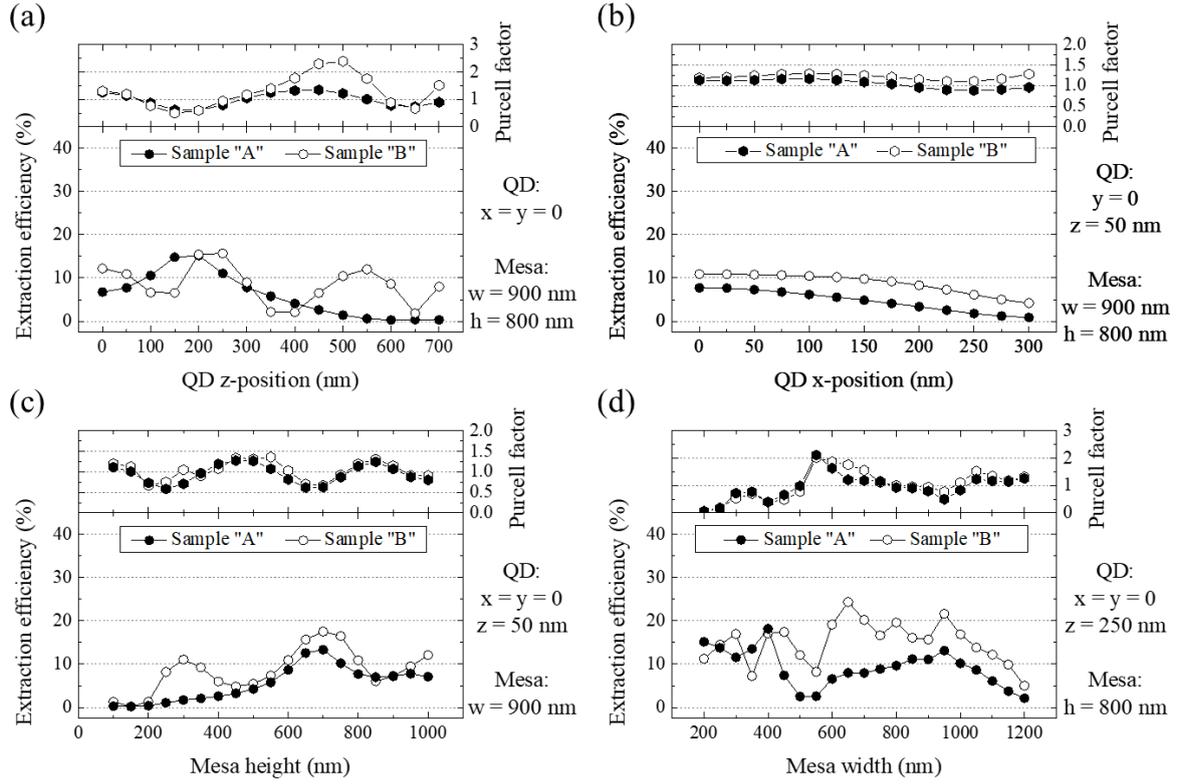

Figure 6. Calculated results of the extraction efficiency and Purcell factor dependence (analogically as in Fig. 5), for the case of cuboid photonic mesa structure, as a function of (a) QD z-position and (b) in-plane x-position, (c) the mesa height and (d) mesa width. The fixed parameters are shown on the right side of the graphs.

*4.4. Extraction efficiency vs. numerical aperture*

We also study the influence of the numerical aperture on the extraction efficiency, keeping in mind that in principle the issue is related to the collection efficiency provided by a "first lens" of the free-space optical setup or by an optical fibre facet placed in the near proximity of the mesa structure. For the former, the NA might be improved by introducing high numerical aperture microscope objectives up to NA = 0.9, which are available on the market, or by using immersion layer in between. Whereas for the latter, the coupling to the optical fibre might be engineered by several strategies, e.g. air-silica microstructured fibres having NA > 0.9 [36]. In Fig. 7(a) we plot the $\eta_{eff}$ versus NA changed from 0.1 to 0.9 for the circular mesa structure using both the "A" and "B" samples of 700 nm in the diameter, the height of 800 nm and the dipole position $z$ = 250 nm. Starting from the initial values obtained already for NA = 0.4, i.e. $\eta_{eff}$ = 6.9% and $\eta_{eff}$ = 21.1%, we notice a significant increase to 11.6% and 31.2% for NA = 0.6 and to 16.5% and 41.7% for NA = 0.9, respectively. In Fig. 7(b) the results are presented for cuboidal mesa structures of width 900 nm, height 800 nm and dipole $z$ position of 250 nm. In this case, we find the maximal extraction efficiencies for NA = 0.9 of $\eta_{eff}$ = 19.2% and $\eta_{eff}$ = 46.9% for the samples "A" and "B", respectively. The highest extraction efficiency obtained for NA = 0.9 is also partly related to simultaneously larger Purcell factor of $F_p$ = 1.77.

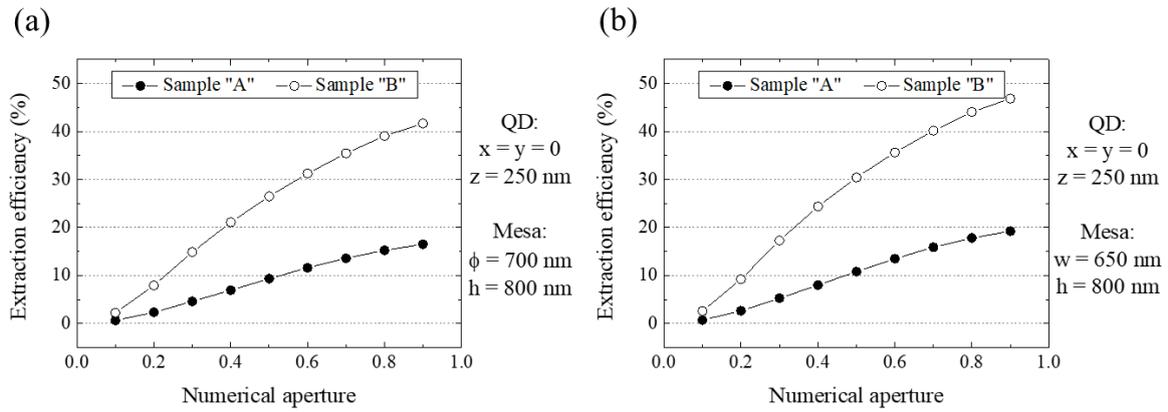

Figure 7. Calculated extraction efficiency for (a) cylindrical and (b) square mesa structure geometry in dependence on the numerical aperture from 0.1 to 0.9 using geometry and dipole position shown on the right side of the graphs.

*4.5. Spectral dependence of the extraction efficiency*

Eventually, we present results on the spectral dependence of the calculated extraction efficiency. In Fig. 8 we consider a detuning in energy (wavelength) in the range of 0.75 -

0.84 eV (1480 - 1650 nm) for a cylindrical mesa structure with a DBR (the sample "B") of 850 nm in diameter, 800 nm in height and the dipole position $z$ = 250 nm, which results in $\eta_{eff}$ = 24.6% and $F_P$ = 1.16 at 1550 nm. We observe that $\eta_{eff}$ > 20% is attained for about 60 nm in the range of 1530 - 1590 nm (which corresponds to 31 meV and the range of 0.78 - 0.81 eV in the energy scale). The Purcell factor for this case changes its value rather insignificantly, in the range of 0.97 – 1.16. In contrast to our observations, the high Q microcavities with the Purcell factors of 5 (and higher) have the full width at half maximum (FWHM) of the cavity mode on the order of 1 meV, i.e. a few nm, or less [37]. In our case, the FWHM of the extraction efficiency spectral dependence is more than 80 nm. Moreover, in our results the Purcell factor is also rather weakly dependent on energy (wavelength) in this range, which shows that the band is mainly defined by the directionality of emission.

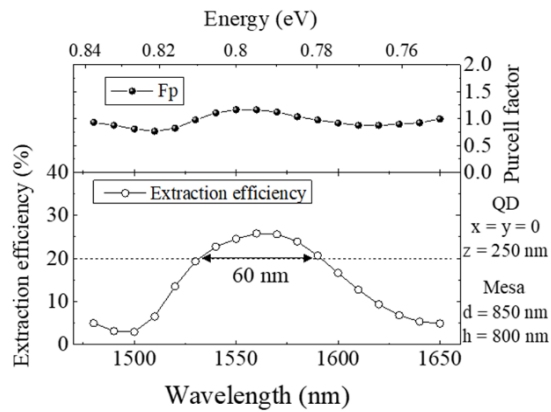

Figure 8. Calculated extraction efficiency for the cylindrical mesa structure with DBR (sample "B") in dependence on energy/wavelength showing extraction efficiency of more than 20% in a range of 31 meV.

## 5. Conclusions

In this work, we examined numerically the extraction efficiency of the quantum-dot-like dipole emission from InP-based photonic mesa structures of circular and square base geometry, suitable for telecommunication range single photon applications at 1.55 µm. We studied the influence of the mesa size and shape and the QD position on the extraction efficiency of emission in two cases: for monolithic InP mesa and for InP mesa on a DBR structure. We found the conditions to reach $\eta_{eff}$ = 18.2% and $\eta_{eff}$ = 24.6%, respectively, for the NA = 0.4 of the collection optics, showing high potential of monolithic InP structure already and even more promising results when the DBR mirror is included. We analysed also the impact of the DBR thickness and how these results depend on the numerical aperture, obtaining the maximum value of $\eta_{eff}$ = 46.9% for NA = 0.9 for a cuboidal mesa. Moreover, in spite of nominally low quality factor of such structures, which on the other hand enables weak spectral sensitivity, we found some influence of the Purcell effect on the dipole emitter which might affect the brightness of such a photonic device, and cannot be neglected in the overall system optimization. The spectral dependence of the extraction efficiency showed that $\eta_{eff}$ > 20% is expected in the range of 60 nm at 1.55 µm. Eventually, it is worth emphasizing that the presented optimization of the InP-based system suitable for the third telecommunication window leads to similar or even better photon extraction efficiency when compared with spectrally broad approaches in more mature GaAs-based structures at shorter wavelengths [12], [23].

**Acknowledgements**

Support by Grant No. 2014/14/M/ST3/00821 of the National Science Centre in Poland

is acknowledged. The authors thank Ania Musiał and Wojtek Rudno-Rudziński for careful reading of the manuscript and many valuable suggestions and comments.